\documentclass[prd,12pt,nofootinbib,preprint]{revtex4}
\usepackage[T1]{fontenc}
\usepackage{amsmath,amssymb}
\usepackage{epsfig}
\usepackage{url}
\usepackage{subfigure}
\usepackage{graphicx}
\usepackage[usenames,dvipsnames]{color}
\usepackage{slashed}
\usepackage{array}
\usepackage{multirow}
\usepackage[colorlinks,citecolor=blue]{hyperref}
\usepackage{color}
\usepackage{cancel}

\def\a {\alpha}
\def\b {\beta}

\def\bar {\overline}

\def\be {\begin{equation}}
\def\ee {\end{equation}}
\def\beq {\begin{equation}}
\def\eeq {\end{equation}}
\def\bea {\begin{eqnarray}}
\def\eea {\end{eqnarray}}

\def\bra {\langle}
\def\ket {\rangle}

\newcommand{\besub}{\begin{subequations}}
\newcommand{\eesub}{\end{subequations}}

\def\beq{\begin{equation}}
\def\eeq{\end{equation}}
\def\barr{\begin{array}}
\def\earr{\end{array}}

\begin{document}
\title{Flavour-alignment in an $S_3$-symmetric Higgs sector and its RG-behaviour}
\author{Nabarun Chakrabarty}
\email{nabarunc@iitk.ac.in, chakrabartynabarun@gmail.com}

\affiliation{Department of Physics, Indian Institute of Technology, Kanpur 208 016, India}
\author{Indrani Chakraborty}
\email{indranic@iitk.ac.in, indrani300888@gmail.com}
\affiliation{Department of Physics, Indian Institute of Technology, Kanpur 208 016, India} 

\begin{abstract}
A three Higgs-doublet model admitting an $S_3$-symmetry can predict the observed pattern of the quark masses and their mixings. However the same symmetry also introduces potential flavour-changing neutral currents at the tree level. We assume in this work that the scalar potential contains appropriate \emph{soft} $S_3$-breaking terms in order to keep the choices of the scalar masses flexible. We identify the parameters in the Yukawa Lagrangian in the quark sector responsible for such FCNCs and constrain them using data from some of the flavour physics observables like meson-decays and meson-mixings. We also validate the corresponding model parameter space with renormalisation group (RG) evaluation. 
\end{abstract} 
\maketitle
\section{Introduction}
\label{intro}

With the recent data from the Large Hadron Collider (LHC) leaning increasingly in favour of the Standard Model (SM), the possibility of additional dynamics beyond the SM however does not fade out. Several issues stemming from both theory and experiments cannot be resolved within the SM alone thereby calling for new physics. One of such issues is the observed pattern of the fermion masses and mixings. While several theoretical scenarios have been put forth to address this issue, a particularly interesting class in this context is based on three Higgs doublets \cite{Aranda:2012bv,Varzielas:2015joa,Ivanov:2012ry,Maniatis:2015kma, Moretti:2015cwa}. The idea here is to connect the three fermionic generations to the three scalar doublets present by means of certain discrete symmetries so as to explain the observed fermion masses and mixings. Discrete symmetries like $A_4, S_3, \Delta_{27}, Z_3$ \cite{Ivanov:2012fp} are a few examples from a longer list that have been embedded in a three Higgs doublet model (3HDM) to the aforementioned effect.

It is not possible to predict the exact number of scalar doublets present in nature from fundamental principles, given that the electroweak $\rho$-parameter does not deviate from unity in presence of doublets alone. In a 
$CP$-conserving 3HDM, one amongst the three $CP$-even scalars must have a mass around 125 GeV in order to comply with Higgs discovery. It is though understood that the couplings of that scalar to fermions and gauge bosons will be scaled with respect to the corresponding SM values, and, the scaling factors will contain mixing angles that connect the gauge basis to the mass eigenstates. However, similar to what is seen in a 2HDM, it is possible to obtain an "alignment-limit" in a 3HDM also, when the couplings coincide with the corresponding SM values. The signal strength data for the 125 GeV scalar is automatically satisfied in that limit. Of course, a 3HDM can be  distinguished from a 2HDM at a collider by virtue of certain cascades of scalars that bear information on the intermediate scalars present. Given that there are no hints of such signals at the LHC, the current scenario allows a 3HDM as much as it allows a 2HDM.

A 3HDM obeying a global $S_3$-symmetry is one such example that permits the desired alignment through its scalar potential. On the other hand, an immediate fallout of an $S_3$-symmetric Yukawa sector is the presence of flavour-changing neutral currents (FCNCs) at the tree-level. The parameters responsible for the same must be typically small in order to satisfy the constraints from meson-mixing and meson-decays. A question then naturally arises that whether such smallness is due to a radiative effect. That is, whether the $S_3$-symmetric Yukawa Lagrangian is a part of a larger symmetry at some high energy scale at which the FCNC parameters vanish, and, following a spontaneous breakdown of the bigger symmetry, they assume appropriately small but non-zero values at the electroweak (EW) scale through evolution under renormalisation group (RG). We have attempted to probe this possibility in this work.

We have computed the one-loop RG equations for all the Yukawa couplings pertaining to the $S_3$-symmetry and identify the ones responsible for FCNC. Without any specific UV-complete theory in mind, we can assume that the FCNC couplings vanish at some scale $\Lambda$. The effective field theory below that scale then corresponds to the $S_3$-symmetric 3HDM. We iterate that our goal is not to make an exhaustive survey of the parameter space of this model taking into account all possible flavour constraints, but, to study to the sensitivity of the FCNC parameters to the aforementioned RG evolution.

The paper is organised as follows : Section \ref{sec2} contains the details of $S_3$-symmetric 3HDM. We present the analysis and results in section \ref{sec3}. Section \ref{sec4} comprises of the discussion of RG-running of Yukawa couplings of the up- and down-sectors. Finally we summarise and conclude in section \ref{sec5}.

\section{The $S_3$-symmetric three Higgs doublet model: Salient features}
\label{sec2}

The $S_3$-symmetric three Higgs doublet model or $S_3$HDM is an extension of the SM based on the discrete group $S_3$, which
comprises three $Y = \frac{1}{2}$ scalar doublets $\phi_1, \phi_2$ and $\phi_3$. 
Of these, $\phi_1$ and $\phi_2$ rotate into each other as doublets under the $S_3$ while $\phi_3$ remains a singlet under the same. The most general scalar potential consistent
with the gauge as well as the $S_3$-symmetry is thus \cite{Das:2015sca,Chakrabarty:2015kmt}
\besub
\bea
V(\phi) &=& \mu_{11}^2(\phi_1^\dagger\phi_1+\phi_2^\dagger\phi_2)+ \mu_{33}^2\phi_3^\dagger\phi_3 \nonumber \\
&& + \lambda_1 (\phi_1^\dagger\phi_1+\phi_2^\dagger\phi_2)^2 +\lambda_2 (\phi_1^\dagger\phi_2 -\phi_2^\dagger\phi_1)^2 +\lambda_3 \left\{(\phi_1^\dagger\phi_2+\phi_2^\dagger\phi_1)^2 +(\phi_1^\dagger\phi_1-\phi_2^\dagger\phi_2) ^2\right\} \nonumber \\
&& +\lambda_4 \left\{(\phi_3^\dagger\phi_1)(\phi_1^\dagger\phi_2+\phi_2^\dagger\phi_1) +(\phi_3^\dagger\phi_2)(\phi_1^\dagger\phi_1-\phi_2^\dagger\phi_2) + {\rm h. c.}\right\} \nonumber \\
&& +\lambda_5(\phi_3^\dagger\phi_3)(\phi_1^\dagger\phi_1+\phi_2^\dagger\phi_2) + \lambda_6 \left\{(\phi_3^\dagger\phi_1)(\phi_1^\dagger\phi_3)+(\phi_3^\dagger\phi_2)(\phi_2^\dagger\phi_3)\right\} \nonumber \\
&& +\lambda_7 \left\{(\phi_3^\dagger\phi_1)(\phi_3^\dagger\phi_1) + (\phi_3^\dagger\phi_2)(\phi_3^\dagger\phi_2) +{\rm h. c.}\right\} +\lambda_8(\phi_3^\dagger\phi_3)^2 \,.
\eea
\eesub

We take all the quartic couplings to be real to forbid $CP$-violation arising from the scalar sector. Following electroweak symmetry breaking (EWSB), the doublets can be expressed as
\bea
\phi_{i} = \frac{1}{\sqrt{2}} \begin{pmatrix}
\sqrt{2} w_i^{+} \\
v_i + h_i + i z_i
\end{pmatrix}~ \rm{for}~\textit{i} = 1, 2, 3.
\label{e:doublet}
\eea
The vacuum expectation values (VEVs) $v_1, v_2, v_3$ satisfy $v_1^2 + v_2^2 + v_3^2 = (246 ~\text{GeV})^2$. In terms of the mass eigenstates, the spectrum consists of three $CP$-even scalars $h, H_1, H_2$, two $CP$-odd scalars $A_1, A_2$, and, two charged scalars 
$H_1^+, H_2^+$. The scalars in the mass eigenbasis are connected to the ones in the gauge eigenbasis through unitary transformations. And the form of such unitary matrices depends on whether the $S_3$-invariance of the scalar potential is exact or allowed to be broken by terms of mass dimension-2. In the case of an exact $S_3$ symmetry, minimising the scalar potential enforces $v_1 = \sqrt{3} v_2$ \cite{Das:2015sca,Chakrabarty:2015kmt} if the conditions obtained thereafter are to be consistent with the $S_3$-invariance. A tan$\beta = \frac{2 v_2}{v_3}$ can be defined in that case similarly as in a 2HDM. It is then seen that the diagonalizing matrices can be parametrised by two mixing angles, i.e., $\a$ and the aforementioned $\b$. Exact forms of the unitary matrices can be seen in \cite{Das:2015sca} and therefore are not shown here for brevity. Similar
to the case of a 2HDM, the relation $\a = \b - \frac{\pi}{2}$ corresponds to the alignment, when the couplings of $h$ to fermions and gauge bosons become equal to their corresponding SM values. Therefore, apart from the radiatively induced $h \rightarrow \gamma \gamma$ channel, the LHC data on the signal strengths of $h$ corresponding to the other channels is automatically satisfied upon going to the $\a = \b - \frac{\pi}{2}$ limit.

The perturbativity and unitarity bounds on the quartic couplings $\lambda_i$'s put an upper bound of $< 1$ TeV on the non-standard masses of the model \cite{Das:2014fea}. To increase the non-standard scalar masses, (later we shall discuss that this is required to satisfy the flavour physics constraints) $S_3$-symmetry is softy broken by dimension-2 operators.
Then, the $CP$-even sector for instance, relates the mass eigenbasis to the gauge eigenbasis through a most general $3 \times 3$ orthogonal matrix $\mathcal{O}$ as follows.

\bea
\begin{pmatrix}
h_1 \\
h_2 \\
h_3 
\end{pmatrix} &=& 
\begin{pmatrix}
O_{11}
& O_{12}
& O_{13} \\
O_{21}
& O_{22}
& O_{23} \\
O_{31}
& O_{32}
& O_{33}
\end{pmatrix} 
\begin{pmatrix}
h \\
H_1 \\
H_2 
\end{pmatrix}
\eea

where,

\bea
O_{11} &=& c_\phi c_\psi - c_\theta s_\phi s_\psi \,, \nonumber \\
O_{12} &=& -c_\phi s_\psi - c_\theta s_\phi c_\psi \,, \nonumber \\
O_{13} &=& s_\phi s_\theta \,, \nonumber \\
O_{21} &=& s_\phi c_\psi + c_\theta c_\phi s_\psi \,, \nonumber \\
O_{22} &=& -s_\phi s_\psi + c_\theta c_\phi c_\psi \,, \nonumber \\
O_{23} &=& -c_\phi s_\theta \,, \nonumber \\
O_{31} &=& s_\psi s_\theta  \,, \nonumber \\
O_{32} &=& c_\psi s_\theta    \,, \nonumber \\
O_{33} &=& c_\theta  \,.
\eea
$\theta, \psi, \phi$ being mixing angles.

Now $S_3$-symmetric most general Yukawa potential for up-type quark sector can be written as \cite{Das:2015sca},

\begin{eqnarray}
-  \mathcal{L}_Y^{u} &=&
\null  y_{1u} \Big( \bar Q_{1} \tilde\phi_3 u_{1R} 
+ \bar Q_2 \tilde\phi_3 u_{2R} \Big)
+ y_{2u} \Big\{ \Big( \bar Q_{1}\tilde\phi_2 + \bar
Q_2\tilde\phi_{1}\Big) u_{1R} + 
\Big( \bar Q_{1}\tilde\phi_{1} -
\bar Q_2\tilde\phi_2 \Big)u_{2R} \Big\} \nonumber \\*
&& \null + y_{3u} \bar Q_3\tilde\phi_3u_{3R}
+y_{4u} \bar Q_3 \Big( \tilde\phi_1 u_{1R} + \tilde\phi_2u_{2R} \Big)
+y_{5u} \Big( 
\bar Q_1 \tilde\phi_1 + \bar Q_2\tilde\phi_2 \Big)
u_{3R} + {\rm h.c.}
\label{uYuk}
\end{eqnarray}

Yukawa Lagrangian for the down-sector can be obtained by replacing $u \rightarrow d$ and $\tilde{\phi} \rightarrow \phi$. It should be noted that the fields $u_i$ and $d_i$ presented here do not denote physical quark fields. Their superpositions which are eigenstates will be given later. Following EWSB, mass matrices for the fermions then have the 
following texture \cite{Das:2015sca}
\bea
{\cal M}_f = \frac{1}{\sqrt 2}\begin{pmatrix}
y_{1f} v_3 + y_{2f} v_2 & y_{2f} v_1 & y_{5f} v_1 \\
y_{2f} v_1  & y_{1f} v_3 - y_{2f} v_2 & y_{5f} v_2 \\
y_{4f} v_1  & y_{4f} v_2 & y_{3f} v_3 \\
\end{pmatrix} \,, \qquad {\rm  with~} f=u,d,l \,.
\label{f-matrix}
\eea

We point out that ${\cal M}_f$ in Eq.(\ref{f-matrix}) is not Hermitian for $y_{4f},y_{5f} \neq 0$ and therefore, is brought to a diagonal form by the following bi-unitary transformation
\besub
\bea
V_L^{\dagger} {\cal M}_f V_R &=& {\rm diag}(m_1,m_2,m_3), \\
m_1 &=& \frac{1}{\sqrt{2}}(y_{1f} v_3 - 2 y_{2f} v_2) \\
m_{3,2} &=& \frac{1}{2 \sqrt{2}}(2 y_{2f} v_2 + (y_{1f} + y_{3f}) v_3 \nonumber \\
&&\pm \sqrt{(y_{1f} v_3 + 2 y_{2f} v_2 - y_{3f} v_3)^2 + 16 y_{4f} y_{5f} v^2_2})
\eea 
\label{diag}
\eesub
where, in Eq.(\ref{diag}), $m_i$ denotes the mass of the $i$th generation fermion. It is
therefore possible to reproduce the observed values of the fermion masses by tuning the
various Yukawa couplings and tan$\beta$ appropriately.

The
matrices $V_L$ and $V_R$ induce flavour-changing couplings with the Higgses in this model. Exact structure of the flavour-conserving as well as flavour-changing couplings can be found in appendix \ref{appendix_couplings}.

\section{Analysis and results}
\label{sec3}
From appendix \ref{appendix_couplings}, it can be seen that the flavour-changing couplings of SM Higgs involving the third generation of fermions are proportional to $y_{5f}$, {\em i.e.} by taking $y_{5f}$ to be negligible, one can ensure small flavour-changing couplings for the SM Higgs. Since the mass matrix of fermions is hermitian for $y_{4f}, y_{5f} = 0$ \footnote{$y_{4f}, y_{5f} = 0$ are attributed to the following global symmetry (in addition to $S_3$),  
\besub
\bea
\phi_{1,2,3} \to \phi_{1,2,3}, \\ 
Q_3,u_3 \to  Q_3,u_3, \\
u_{1,2} \to e^{i\theta} u_{1,2}, \\
Q_{1,2} \to e^{i\theta} Q_{1,2}.
\eea
\eesub 
The terms in the Yukawa Lagrangian with the coefficients $y_{1f},y_{2f},y_{3f}$ are invariant under the above symmetry while the terms with the coefficients $y_{4f},y_{5f}$ break it.
This symmetry ensures that $y_{4f},y_{5f}$ (with $f=u,d$) are radiatively protected. }, we assume $y_{4f}, y_{5f}$ to be tiny for the entire analysis, which in turn makes the flavour-changing couplings to SM Higgs small.

Neglecting the tiny $y_{4f}$ and $y_{5f}$, the rest of the three flavour-changing Yukawa couplings $y_{1f}$, $y_{2f}$ and $y_{3f}$ are fixed by the fermion masses $m_1$, $m_2$ and $m_3$ as mentioned below :
\besub
\bea
y_{1f} &\simeq & \frac{(m_1 + m_2)}{\sqrt{2} v_3} , \\
y_{2f} &\simeq & \frac{(m_2 - m_1)}{2 \sqrt{2} v_2} , \\
y_{3f} &\simeq & \frac{\sqrt{2} m_3}{v_3} .
\eea
\label{eq1}
\eesub
For analysis, we have varied $y_{4f}$ and $y_{5f}$ as,
\bea
-0.005 \leq y_{4f} \leq 0.005 , ~ -0.005 \leq y_{5f} \leq 0.005 .
\eea

$v_1, v_2 $ and $v_3$ can be expressed in terms of the mixing-angles $\beta$ and $\gamma$ as,
\besub
\bea
v_1 &= & v~ \rm{sin} \beta ~ \rm{cos} \gamma , \\
v_2 &= & v~ \rm{sin} \beta ~ \rm{sin} \gamma , \\
v_3 &= & v~ \rm{cos} \beta .
\eea
\eesub

We have used the masses of the mass eigenstates as,
\bea
m_h = 125.3~ {\rm{GeV}} , m_{H_1} = m_{H_2} = m_{A_1} = m_{A_2} = 1 ~{\rm{TeV}}.
\eea


%
%
%

To ensure that the lightest Higgs ($h$) of the model behaves as SM Higgs, the couplings of $h$ to gauge bosons as well as fermions (mentioned in appendix \ref{appendix_couplings}), are considered to be identical to that of the SM-Higgs by suitable choices of the angles $\beta, \gamma, \theta, \phi, \psi$. While fixing $\gamma$, we have taken the flavour-changing couplings of $h$ to first two generations of up type and down type quarks, {\em i.e.} $y_{huc}$ and $y_{hds}$ to be zero.
\begin{table}
\begin{center}
\begin{tabular}{ |c|c|c|c| } 
\hline \hline
Benchmark & Angle & $y_{iu}$ & $y_{id}$ \\
\hline
BP1 & $\beta =$ 0.314159 & $y_{1u} = 0.00385$ & $y_{1d} = 0.00030$ \\ 
& $\gamma =$ 0.839897 & $y_{2u} = 0.00794$ & $y_{2d} = 0.00056$   \\ 
&  $\theta =$ 1.20 & $y_{3u} = 0.99708$ & $y_{3d} = 0.01872$   \\ 
& $\phi =$ 4.94 &  &   \\
& $\psi =$ 1.82 &  &   \\
\hline \hline

BP2 & $\beta =$ 0.314159 & $y_{1u} = 0.00385$ & $y_{1d} = 0.00030$  \\ 
& $\gamma =$ 1.12824 & $y_{2u} = 0.00654$ & $y_{2d} = 0.00046$   \\ 
&  $\theta =$ 2.10 & $y_{3u} = 0.99708$ & $y_{3d} = 0.01872$   \\ 
& $\phi =$ 2.54 &  &   \\
& $\psi =$ 1.49 &  &   \\
\hline \hline
\end{tabular}
\caption{The angles and the values of the Yukawa couplings $y_{iu}$, $y_{id}$ (for $i = 1,2,3$) at the electroweak scale are given for BP1 and BP2. }
\label{table1}
\end{center}
\end{table}
Thus two Benchmark points are chosen with different values of mixing angles as shown in Table \ref{table1}. The values of $y_{1f}, y_{2f}, y_{3f}$ at the electroweak scale are fixed by Eq.(\ref{eq1}) are given in Table \ref{table1} for two different benchmark points BP1 and BP2. We have taken $y_{4u}$ and $y_{5u}$ to be zero at the electroweak scale. The corresponding values for $y_{4d}$ and $y_{5d}$ ( $<< y_{1d}, y_{2d}, y_{3d}$  ) at the electroweak scale are fixed by the flavour physics constraints like meson-mixing, meson-decays etc. as described in the next subsection. In Figure \ref{figure1}, the cyan colored points represent the parameter space spanned by $y_{4d}$ and $y_{5d}$ at electroweak scale for two different benchmark points.

\subsection{Flavour Physics constraints}

In this subsection, we discuss the relevant processes contributing to flavour physics constraints on the flavour-changing couplings to the fermions.

\subsubsection{$B_s \rightarrow \mu^+ \mu^- $} 
\label{subsub1}
The effective Hamiltonian for the process $B_s \rightarrow \mu^+ \mu^- $ can be calculated as \cite{Buchalla:1995vs},
\bea
\mathcal{H}_{\rm{eff}} = - \frac{G_F}{\sqrt{2}} \frac{\alpha_{\rm{em}}}{\pi s_W^2} V_{tb} V_{ts}^*(C_A \mathcal{O}_A + C_S \mathcal{O}_S + C_P \mathcal{O}_P + C_S' \mathcal{O}_S' + C_P' \mathcal{O}_P') + {\rm h.c.} \,
\label{lgng1}
\eea

where $G_F$ is the Fermi constant, $\alpha_{\rm{em}}$ is  the fine structure constant, $V_{ij}$ are the Cabibbo-Kobayashi-Masakawa (CKM) matrix elements and $s_W = {\rm {sin}} \theta_W, \theta_W$ being the Weinberg angle.

The operators $\mathcal{O}_i$ and $\mathcal{O}_i'$ are defined as,
\bea
\mathcal{O}_A &=& (\bar{s}\gamma_\mu P_L b)(\bar{\mu} \gamma^\mu \gamma_5 \mu)\,, \\
\mathcal{O}_S &=& (\bar{s} P_R b)(\bar{\mu} \mu) \,, \\
\mathcal{O}_P &=& (\bar{s} P_R b)(\bar{\mu} \gamma_5 \mu) \,, \\
\mathcal{O}_S' &=& (\bar{s} P_L b)(\bar{\mu} \mu) \,, \\
\mathcal{O}_P' &=& (\bar{s} P_L b)(\bar{\mu} \gamma_5 \mu) \,.
\eea

Here the Wilson coefficient $C_A$ receives contribution from Standard model only. Where as, within the scope of Standard model, the Wilson coefficients $C_S^{\rm{SM}}, C_S^{'\rm{SM}}, C_P^{\rm{SM}}, C_P^{'\rm{SM}}$ coming from the Higgs-penguin diagrams are highly suppressed. 

That is why we have approximated,
\bea
C_S^{\rm{SM}} = C_S^{'\rm{SM}} = C_P^{\rm{SM}} = C_P^{'\rm{SM}} = 0\,.
\eea
The New physics (NP) contributions to the scalar and pseudoscalar Wilson coefficients are,
\bea
C_S^{\rm{NP}} &=& - \kappa \sum_{\Phi_S}(\frac{y_{\Phi_S sb}~ y_{\Phi_S \mu\mu}}{m_{\Phi_S}^2}),~ \Phi_S = h, H_1, H_2\,. \\
C_S^{'\rm{NP}} &=& C_S^{\rm{NP}} \,, \\
C_P^{\rm{NP}} &=&  \kappa \sum_{\Phi_P}(\frac{y_{\Phi_P sb}~ y_{\Phi_P \mu\mu}}{m_{\Phi_P}^2}),~ \Phi_P =  A_1, A_2\,. \\
C_P^{'\rm{NP}} &=& - C_P^{\rm{NP}} \,, 
\eea
with $\kappa = \frac{\pi^2}{G_F^2 m_W^2 V_{tb} V_{ts}^*}$, $m_W$ being mass of $W$-boson. Here $y_{\Phi_{S(P)} sb}$ is the Yukawa coupling between scalar (pseudoscalar) and first two generations of down quarks and $y_{\Phi_{S(P)} \mu \mu}$ is the Yukawa coupling between scalar (pseudoscalar) and muons.

From the Hamiltonian in eq.(\ref{lgng1}) the branching ratio of the process $B_s \rightarrow \mu^+ \mu^-$ is \cite{Li:2014fea,Cheng:2015yfu},
\bea
{\rm{Br}}(B_s \rightarrow \mu^+ \mu^- ) = \frac{\tau_{B_s} G_F^4 m_W^4}{8 \pi^5} |V_{tb} V_{ts}^*|^2 f_{B_s}^2 m_{B_s} m_\mu^2 \sqrt{ 1 - \frac{4 m_\mu^2}{m_{B_s}^2}} (|P|^2 + |S|^2)\,.
\eea
where $m_{B_s}$, $\tau_{B_s}$ and $f_{B_s}$ are the mass, lifetime and decay constant of the $B_s$ meson respectively (values can be found in reference \cite{Tanabashi:2018oca}) and
\bea
P &\equiv & C_A + \frac{m_{B_s}^2}{2 m_\mu} \left(\frac{m_b}{m_b + m_s}\right) (C_P - C_P')\,, \nonumber \\
S & \equiv & \sqrt{ 1 - \frac{4 m_\mu^2}{m_{B_s}^2}} \frac{m_{B_s}^2}{2 m_\mu} \left(\frac{m_b}{m_b + m_s}\right) (C_S - C_S')\,,
\eea
where $ C_A = - \eta_Y Y_0$ , $\eta_Y = 1.0113$ and 
$Y_0 = \frac{x}{8} \left(\frac{(4-x)}{(1-x)} + \frac{3 x ~{\rm{ln}}x}{(1-x)^2}\right)$, 
$x = \frac{m_t^2}{m_W^2}$ \cite{Buras:2012ru}, $m_t$, $m_b$, $m_s$ and $m_\mu$ are top quark , bottom quark and strange quark masses and muon mass respectively.

For $B_s-\bar{B}_s$ oscillations, the measured branching ratio of $B_s \rightarrow \mu^+ \mu^-$ should be calculated as time-integrated one \cite{DeBruyn:2012wk},

\bea
\bar{\mathcal{B}}(B_s \rightarrow \mu^+ \mu^-) = \left(\frac{1 + \mathcal{A}_{\Delta \Gamma} y_s}{1 - y_s^2}\right) {\rm{Br}}(B_s \rightarrow \mu^+ \mu^-) \,.
\eea 
where
\bea
y_s &=& \frac{\Gamma_s^L - \Gamma_s^H}{\Gamma_s^L + \Gamma_s^H} = \frac{\Delta \Gamma_s}{2 \Gamma_s} \,, \nonumber  \\
\mathcal{A}_{\Delta \Gamma} &=& \frac{|P|^2 {\rm{cos}} (2 \phi_P - \phi_s^{NP}) - |S|^2 {\rm{cos}} (2 \phi_S - \phi_s^{NP}) }{|P|^2 + |S|^2} \,.
\eea
Here $\phi_{S(P)}$ are the phases associated with $S(P)$, $\phi_s^{NP}$ is the CP phase coming from $B_s-\bar{B}_s$ mixing. Within the scope of Standard model, $\mathcal{A}_{\Delta \Gamma} = 1$. $\Gamma_s^L$ and $\Gamma_s^H$ are the decay widths of the light and heavy mass eigenstates of $B_s$.

Since the couplings $y_{\Phi_{S(P)} sb}$ and $y_{\Phi_{S(P)} \mu \mu}$ are constrained by the $\bar{\mathcal{B}}(B_s \rightarrow \mu^+ \mu^-)$ data, from appendix \ref{appendix_couplings}, this is obvious that stringent bounds are imposed on the mixing angles and some of the Yukawa couplings in the down-sector. 

During the analysis, we have used $2\sigma$-experimental value of $\bar{\mathcal{B}}(B_s \rightarrow \mu^+ \mu^-)$ (available in Table \ref{table2}) for data fitting.

\subsubsection{$B_d \rightarrow \mu^+ \mu^- $}
All formulae are same as in the case of $B_s \rightarrow \mu^+ \mu^-$ in subsection \ref{subsub1}, after the replacement $s \rightarrow d$. Here also we have used the experimental bound on the branching ratio (quoted in Table \ref{table2}) within $2 \sigma$-window.
\subsubsection{$B_q-\bar{B}_q$ mixing, $q = s,d$}

The effective Hamiltonian for $B_s-\bar{B}_s$-mixing can be written as \cite{Buras:2001ra,Zhang:2018nmy},
\bea
\mathcal{H}_{\rm eff}^{\Delta B = 2} = \frac{G_F^2}{16 \pi^2} m_W^2 (V_{tb} V_{tq}^*)^2 \sum_i C_i \mathcal{O}_i + {\rm h.c.} \,,
\eea
where the operators $\mathcal{O}_i$ can be expressed as \cite{Buras:2001ra,Zhang:2018nmy},
\bea
\mathcal{O}^{VLL}_1 &=& (\bar{q}^\alpha \gamma_\mu P_L b^\alpha)(\bar{q}^\beta \gamma_\mu P_L b^\beta)\,, \nonumber \\
\mathcal{O}^{SLL}_1 &=& (\bar{q}^\alpha P_L b^\alpha)(\bar{q}^\beta P_L b^\beta) \,, \nonumber \\
\mathcal{O}^{SRR}_1 &=& (\bar{q}^\alpha P_R b^\alpha)(\bar{q}^\beta P_R b^\beta) \,, \nonumber \\
\mathcal{O}^{LR}_2 &=& (\bar{q}^\alpha P_L b^\alpha)(\bar{q}^\beta P_R b^\beta) \,
\eea
$\alpha$ and $\beta$ being the colour indices (not to be confused with mixing angles).

 The contribution from the Standard model comes via $\mathcal{O}^{VLL}_1$. The Standard model contribution to the transition matrix element of $B_q-\bar{B}_q$ mixing is given by \cite{Buras:2001ra,Zhang:2018nmy},
\bea
M_{12}^{q(SM)} &=&  \frac{G_F^2}{16 \pi^2} m_W^2 (V_{tb} V_{tq}^*)^2 \left[C^{VLL}_1 \bra \mathcal{O}^{VLL}_1 \ket \right]\,, \nonumber \\
&=& \frac{G_F^2 m_W^2 m_{B_q}}{12 \pi^2} S_0(x_t)\eta_{2B} |V_{tq}^* V_{tb}|^2 f_{B_q}^2 \hat{B}_{B_q}^{(1)} \,, \nonumber \\
\eea
where, 
\bea
S_0(x_t) &=& \frac{4 x_t - 11 x_t^2 + x_t^3}{4(1-x_t)^2} - \frac{3 x_t^3~ {\rm ln} x_t}{2 (1 - x_t)^3}\,, \nonumber \\
x_t &=& \frac{m_t^2 (\mu_t)}{m_W^2} \,, \nonumber \\
\eta_{2B} &=& \left[\alpha_s (\mu_W)\right]^{\frac{6}{23}} \,, \nonumber \\
\hat{B}_{B_q}^{(1)} &=& 1.4 \,
\eea

The NP-contributions reflect through the rest of the operators $\mathcal{O}^{SLL}_1$, $\mathcal{O}^{SRR}_1$, $\mathcal{O}^{LR}_2$ generated by Higgs flavour-changing neutral current (FCNC) interactions. The corresponding Wilson coefficients contain the model informations and are calculated as,

\bea
C^{SRR}_1 &=& \frac{16 \pi^2}{G_F^2 m_W^2 (V_{tb} V_{tq}^*)^2 }\left[ \sum_{\Phi_S} \frac{y_{\Phi_S b q}^2}{m_{\Phi_S}^2} - \sum_{\Phi_P} \frac{y_{\Phi_P b q}^2}{m_{\Phi_P}^2}\right] \,, \nonumber \\
C^{SLL}_1 &=& C^{SRR}_1 \,, \nonumber \\
C^{LR}_2 &=& \frac{32 \pi^2}{G_F^2 m_W^2 (V_{tb} V_{tq}^*)^2 }\left[ \sum_{\Phi_S} \frac{y_{\Phi_S b q}^2}{m_{\Phi_S}^2} + \sum_{\Phi_P} \frac{y_{\Phi_P b q}^2}{m_{\Phi_P}^2}\right] \,.
\label{eqn2}
\eea
where, $\Phi_S = h, H_1, H_2$ and $\Phi_P = A_1, A_2$.

Overall transition matrix element of $B_q-\bar{B}_q$ mixing containing Standard model and NP contribution, is given by \cite{Buras:2001ra,Zhang:2018nmy},
\bea
M_{12}^q &=& \bra B_q |\mathcal{H}_{\rm eff}^{\Delta B = 2}|\bar{B}_q \ket \,, \nonumber \\
&=& \frac{G_F^2}{16 \pi^2} m_W^2 (V_{tb} V_{tq}^*)^2 \sum_i C_i  \bra B_q |\mathcal{O}_i|\bar{B}_q \ket \,. \nonumber \\
&=& M_{12}^{q(SM)} + M_{12}^{q(NP)} \,, \nonumber \\
&=& M_{12}^{q(SM)} + \frac{G_F^2}{16 \pi^2} m_W^2 (V_{tb} V_{tq}^*)^2 \left[C^{SLL, NP}_1 \bra \mathcal{O}^{SLL}_1 \ket + C^{SRR, NP}_1 \bra   \mathcal{O}^{SRR}_1 \ket + C^{LR, NP}_2 \bra \mathcal{O}^{LR}_2  \ket \right] \,. \nonumber \\
\eea
with \cite{Bazavov:2016nty},

\bea
\bra \mathcal{O}^{VLL}_1 \ket &=& c_1 f_{B_q}^2 m_{B_q}^2 B_{B_q}^{(1)} (\mu)\,, \nonumber \\
\bra \mathcal{O}^{SLL}_1 \ket &=& c_2 \left(\frac{m_{B_q}}{m_b(\mu) + m_q(\mu)}\right)^2 f_{B_q}^2 m_{B_q}^2 B_{B_q}^{(2)} (\mu) \,, \nonumber \\
\bra \mathcal{O}^{SRR}_1 \ket &=& \bra \mathcal{O}^{SLL}_1 \ket \,, \nonumber \\
\bra \mathcal{O}^{LR}_2  \ket &=& c_4 \left[\left(\frac{m_{B_q}}{m_b(\mu) + m_q(\mu)}\right)^2 + d_4 \right] f_{B_q}^2 m_{B_q}^2 B_{B_q}^{(4)} (\mu) \,, \eea
where $c_1 = \frac{2}{3}, ~ c_2 = - \frac{5}{12} , ~ c_4 = \frac{1}{2}, d_4 = \frac{1}{6} , B_{B_q}^{(1,2,4)}(\mu) = 1$. $f_{B_q}$, $m_{B_q}$ can be found in \cite{FermilabLattice:2016ipl,Gabbiani:1996hi}.
 
Now the mass difference between $B_q-\bar{B}_q$ can be written as,
\bea
\Delta m_q &=& 2 |M_{12}^q| \,.
\eea
Since all the Yukawa couplings are taken to be real, the CP-violation phase becomes zero.

From Eq.(\ref{eqn2}), it is evident that the mass difference $\Delta m_q$ is solely dependent on Yukawa couplings $y_{\Phi_{S(P)}bq}$ and masses $m_{\Phi_{S(P)}}$. The experimental constraint on $\Delta m_q$ can be translated to some bound on the mixing angles and some of the Yukawa couplings in the down-sector. Here also we have used $2\sigma$- experimental values of $\Delta m_q$ available in Table \ref{table2}.

\subsubsection{$K_0-\bar{K}_0$ mixing}

For brevity, we do not write detailed formulae for $K_0-\bar{K}_0$ mixing, which are much similar to $B_q-\bar{B}_q$ oscillations. The detailed formulae for $K_0-\bar{K}_0$ mixing can be found in reference \cite{Buras:2001ra, Ciuchini:1998ix}. 

The NP contribution to the mass difference $\Delta m_K$ involves the Yukawa couplings $y_{\Phi_{S(P)} ds}$ and masses $m_{\Phi_{S(P)}}$. They will restrict the mixing angles and Yukawa couplings in turn.

The hadronic uncertainties in $K_0-\bar{K}_0$ mixing being relatively large \cite{Buras:2013ooa, Bertolini:2014sua}, we allow for 50$\%$ range of $(\Delta m_K)_{exp}$ (can be found in Table \ref{table2}), while considering the Higgs FCNC effects to $\Delta m_K$. For this conservative estimate, we have followed \cite{Bertolini:2014sua}.

The aforementioned relevant flavour physics observables are tabulated in Table \ref{table2}.
\subsubsection{$D_0-\bar{D}_0$ mixing and $t \rightarrow c h$}
The constraints on the flavour-changing Yukawa couplings in the up-sector comes from $D_0-\bar{D}_0$ mixing and the process $t \rightarrow c h$. $D_0-\bar{D}_0$ mixing imposes constraints on couplings $y_{\Phi_{S(P)} uc}$, similar to $B_q-\bar{B}_q $ and $K_0-\bar{K}_0$ mixing in the down-sector. Since $y_{\Phi_{S(P)} uc}$ is proportional to $y_{2u}$ which is fixed by the quark masses, the mixing angles are only affected by this constraint. Detailed formulae can be found in reference \cite{Lunghi:2007ak}. We have used $2 \sigma $-allowed range of the experimental value for the mass difference $\Delta m_{D_0-\bar{D}_0}$ (mentioned in Table \ref{table2}).

The process $t \rightarrow c h$ gives a bound on the flavour-changing coupling $y_{hct}$ \cite{Aaboud:2017mfd}, which is somehow less stringent.
\begin{table}
\begin{center}
\begin{tabular}{| c | c | c |}
\hline
Observables & SM value & Experimental value \\
\hline
${\cal{\bar{B}}}(B_s \rightarrow \mu^+ \mu^-)$(10$^{-9})$ & 3.66 $\pm 0.14$ \cite{LHCb:2021vsc}& 3.09 $^{+0.46~+0.15}_{-0.43~-0.11}$ \cite{LHCb:2021vsc}\\ 
\hline
$Br(B_d \rightarrow \mu^+ \mu^-)$(10$^{-10})$ & 1.03 $\pm 0.05$ \cite{LHCb:2021vsc} & 1.2$^{+0.8}_{-0.7}\pm 0.1$ \cite{LHCb:2021vsc} \\
\hline
$\Delta m_s$ (ps$^{-1}$) & 18.3$\pm2.7$ \cite{Jubb:2016mvq,HFLAV:2016hnz} & 17.749$\pm 0.019~({\rm stat})\pm 0.007~({\rm syst.})$ \cite{CDF:2006imy,LHCb:2014iah,LHCb:2019nin,LHCb:2011vae,LHCb:2013fep,LHCb:2013lrq}\\
\hline
$\Delta m_d$ (ps$^{-1}$) & 0.528$\pm 0.078$ \cite{Jubb:2016mvq,HFLAV:2016hnz} & 0.5065$\pm 0.0019$ \cite{HFLAV:2019otj}\\
\hline
$\Delta m_K$ ($10^{-3}$ps$^{-1}$) & 4.68$\pm 1.88$ & 5.293$\pm 0.009$ \cite{Tanabashi:2018oca}\\
\hline \hline
\end{tabular}
\caption{Standard model prediction and experimental values of different flavour physics observables }
\label{table2}
\end{center}
\end{table}
\section{RG-running : bottom-up vs. top-down approach}
\label{sec4}
After imposing aforementioned flavour physics constraints, we have obtained the parameter space spanned by $y_{iu}$ and $y_{id}$ ( $i = 5$) at the electroweak scale. Now one can compute Renormalisation Group Equations (RGEs) of $y_{iu(d)}$ using quark mass matrix in Eq.(\ref{f-matrix}). It should be noted from the RGEs in appendix \ref{RG_EQ}, that RGE for each Yukawa coupling is dependent on both up-type and down-type Yukawa couplings. RGEs for up-type Yukawa couplings can be derived by replacing $d \leftrightarrow u$ in the RGEs of down-type Yukawa couplings.
\subsection{Bottom-up approach}
In the bottom-up approach, we start from the values of $y_{iu(d)}$ at the electroweak scale, keeping $y_{4u} = y_{5u} = 0$ and study the evolution of the couplings under RGEs upto the scale $\Lambda = 10^5 , 10^{11}, 10^{16}$ GeV. 

At electroweak scale, $y_{1u(d)}, y_{2u(d)}, y_{3u(d)}$ are fixed by the masses of the quark and mixing angles. Therefore for a fixed benchmark point, the initial values of these couplings remain same at electroweak scale depending on the mixing angles. But since RGEs of these six couplings also depend on $y_{4u(d)}, y_{5u(d)}$, which decrease with increasing energy scale, $y_{1u(d)}, y_{2u(d)}, y_{3u(d)}$ show similar trend of decreasing with increase of energy scale. 

\begin{figure}[htpb]
\hfill
\subfigure{\includegraphics[width=8cm]{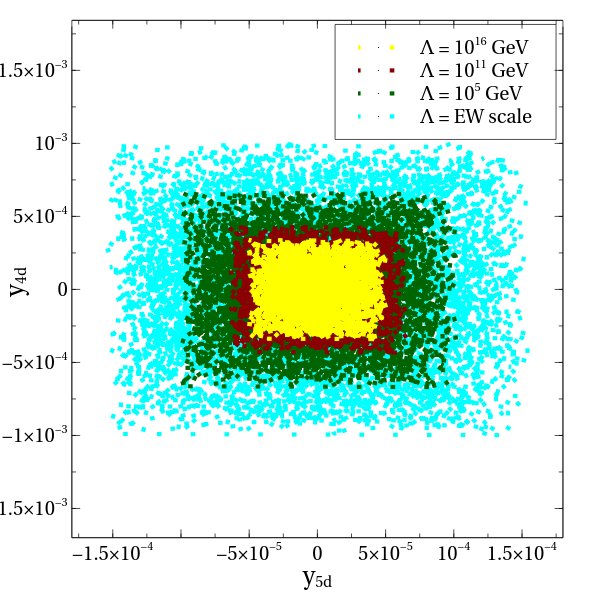}}
\hfill
\subfigure{\includegraphics[width=8cm]{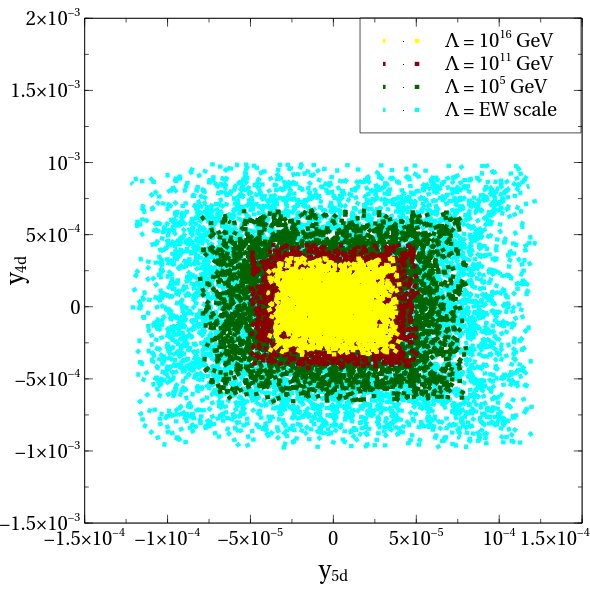}}
\hfill
\caption{Parameter space spanned by $y_{4d}$, $y_{5d}$ for four  different validity scales $\Lambda = {\rm{EW-scale}}, 10^5, 10^{11}, 10^{16}$ GeV. Colour coding is expressed in legends.}
\label{figure1}
\end{figure}

Fig.\ref{figure1} shows that increase in the validity scale $\Lambda$, constraints the allowed parameter space in $y_{4d}-y_{5d}$ plane. Considering the validity of the flavour physics constraints to be the preliminary criteria in the choice of parameters at the EW-scale, one can conclude that the parameter space in the $y_{4d}-y_{5d}$ plane shrinks as the scale of validity increases. One must note that for appropriately small values of $y_{4f}$ and $y_{5f}$ as demanded by the FCNC constraints, the RG evolution of the same does not majorly depend on that of $y_{1f},y_{2f},y_{3f}$. This is apparent from the fact that $\beta$-functions for $y_{4u},y_{4d},y_{5u},y_{5d}$ vanish when 
$y_{4u}=y_{4d}=y_{5u}=y_{5d}$ = 0. This is therefore a fixed point of this theory. Thus the allowed parameter regions in the left and right panels are not majorly different. There are however small differences as can be found upon a careful inspection.

\subsection{Top-down approach}
\begin{figure}[htpb]
\hfill
\subfigure{\includegraphics[width=5cm]{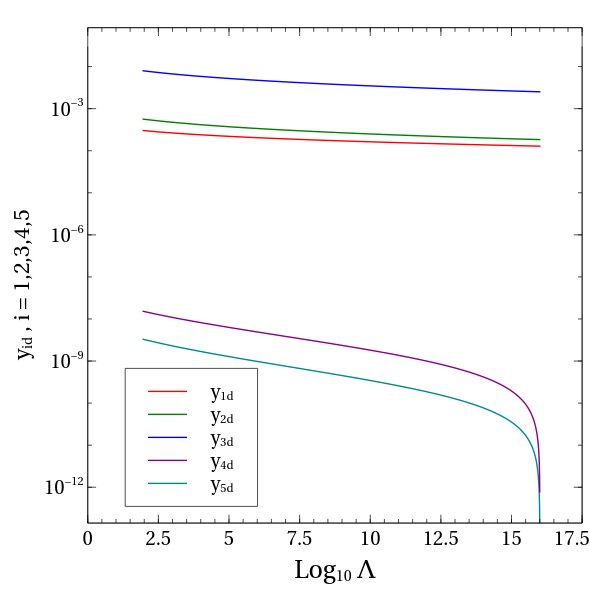}}
\hfill
\subfigure{\includegraphics[width=5cm]{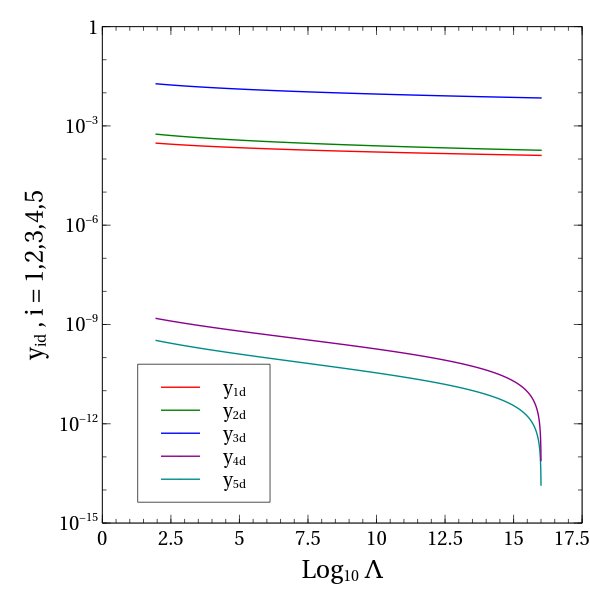}}
\hfill
\subfigure{\includegraphics[width=5cm]{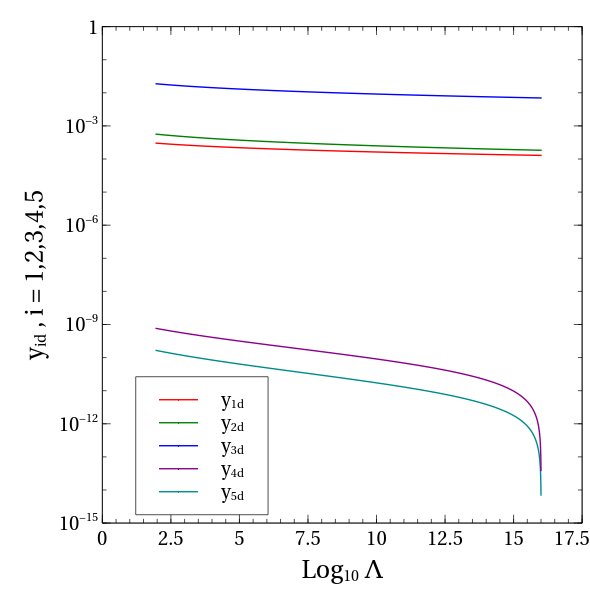}}
\hfill
\subfigure{\includegraphics[width=5cm]{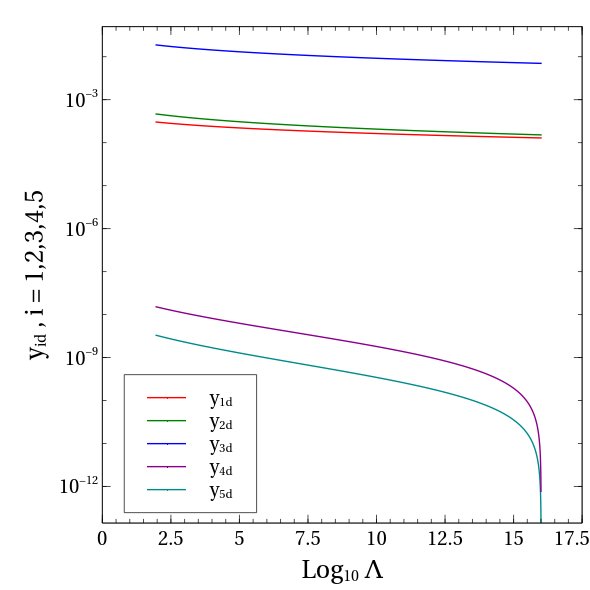}}
\hfill
\subfigure{\includegraphics[width=5cm]{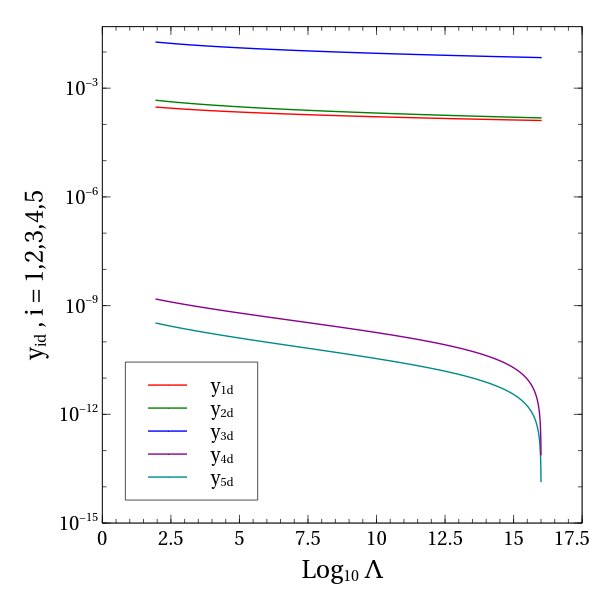}}
\hfill
\subfigure{\includegraphics[width=5cm]{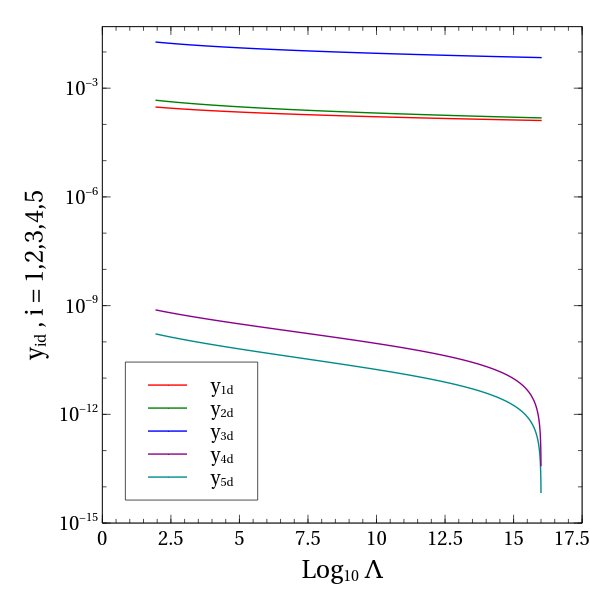}}
\hfill
\caption{Upper panel : $y_{id}$ vs. ${\rm Log}_{10}\Lambda$ plot for BP1 with three different initial values of $y_{4u}, y_{5u}$ at $10^{16}$ GeV. Lower panel : $y_{id}$ vs. ${\rm Log}_{10}\Lambda$ plot for BP2 with three different initial values of $y_{4u}, y_{5u}$ at $10^{16}$ GeV.}
\label{figure2}
\end{figure}

In this section, we consider a reverse-running of all the Yukawa couplings ($y_{iu(d)} , i = 5$) from a higher scale, {\em i.e.} $10^{16}$ GeV to the EW scale and check whether the flavour physics constraints are satisfied at the EW scale or not. From Fig.\ref{figure2} we can find that for each benchmark points (BP1 and BP2), there are three different plots in "$y_{id}$ vs. Log$_{10}\Lambda$" plane, for three different starting values of $y_{4u}$ and $y_{5u}$ ({\em {i.e.}} $10^{-4}, 10^{-5}$ and $5 \times 10^{-6}$) at $10^{16}$ GeV. Corresponding values of $y_{4d}$ and $y_{5d}$ are zero to start with at $10^{16}$ GeV, which might be an artifact of some unknown symmetry.

As we lower the energy scale, since the RG equations are coupled mutually, $y_{4d}, y_{5d}$ can pick up a non-zero but still very small value, which are compatible with flavour physics constraints at the EW scale. The trend of evolution of other Yukawa couplings are same as in the bottom-up approach, {\em i.e.} lower is the energy scale, higher are the Yukawa couplings. Again, the RG evolution curves corresponding to BP1 and BP2 are not appreciably different due to the reason elaborated before.
\section{Conclusion}
\label{sec5}
We have considered the tree level flavour-changing neutral currents in the quark sector of $S_3$-symmetric 3HDM. The flavour-changing Yukawa couplings have been constrained using  perturbativity criteria as well as relevant flavour physics observables coming from meson-decay, meson-mixing etc. in the up-type and down-type quark sector. It can be inferred that the constraints coming from meson mixing put more stringent bound on the flavour-changing couplings compare to the others.

Initially we found a parameter space compatible with the recent flavour physics data, spanned by several flavour-changing Yukawa couplings and mixing angles at the EW scale. Later we have evolved the couplings from EW scale via bottom-up approach, through coupled RG equations to analyse the high scale validity of the model. The trend of evolution of all the Yukawa couplings are similar, {\em i.e.} with increase in energy scale the couplings  decrease.

Finally we have started with zero values of $y_{4d}, y_{5d}$ at $10^{16}$ GeV, as an artifact of some hidden symmetry and evolved them to EW scale via reverse running. We end up with non-zero but negligible values of $y_{4d}, y_{5d}$ generated radiatively at the EW scale, which are still compatible with all the flavour physics constraints. 
\section{Acknowledgements}
NC acknowledges support from DST, India, under Grant Number IFA19-PH237 (INSPIRE Faculty Award). IC acknowledges support from DST, India, under grant number IFA18-PH214 (INSPIRE Faculty Award). 
\appendix

\section{One-loop RG equations}
\label{RG_EQ}

The one-loop beta RG equations for the Yukawa couplings are listed below:
\bea
16 \pi^2 \frac{d y_{1u}}{dt} &=& \frac{1}{2} (9 y_{1d}^2 y_{1u}-8 y_{1d} y_{2d} y_{2u}+4 y_{1l}^2 y_{1u}+15 y_{1u}^3+2 y_{1u} y_{2d}^2+6 y_{1u} y_{2u}^2+6 y_{1u} y_{3d}^2 \nonumber \\
&&+2 y_{1u}
   y_{3l}^2+6 y_{1u} y_{3u}^2+2 y_{1u} y_{4u}^2+y_{1u} y_{5d}^2+y_{1u} y_{5u}^2-4 y_{3d} y_{4u} y_{5d}) + a_u y_{1u} \,, \nonumber \\
16 \pi^2 \frac{d y_{2u}}{dt} &=& \frac{1}{2} (y_{1d}^2 y_{2u}-4 y_{1d} y_{1u} y_{2d}+3 y_{1u}^2 y_{2u}+14 y_{2d}^2 y_{2u}-4 y_{2d} y_{4d} y_{4u}+4 y_{2l}^2 y_{2u}+18 y_{2u}^3 \nonumber \\
&&+6 y_{2u} y_{4d}^2  +2 y_{2u} y_{4l}^2+8 y_{2u} y_{4u}^2+3 y_{2u} y_{5d}^2+2 y_{2u} y_{5l}^2+7 y_{2u} y_{5u}^2) + a_u y_{2u}\,, \nonumber \\
16 \pi^2 \frac{d y_{3u}}{dt} &=& 6 y_{1d}^2 y_{3u}-4 y_{1d} y_{4d} y_{5u}+\frac{1}{2} y_{3u} (4 y_{1l}^2+12 y_{1u}^2+3 y_{3d}^2+2 y_{3l}^2+9 y_{3u}^2 \nonumber \\
&&+2 (y_{4d}^2+y_{4u}^2+2
   y_{5u}^2)) + a_u y_{3u}\,, \nonumber \\
16 \pi^2 \frac{d y_{_{4u}}}{dt} &=& y_{1u}^2 y_{4u}-2 y_{1u} y_{3d} y_{5d}+6 y_{2d}^2 y_{4u}-4 y_{2d} y_{2u} y_{4d}+\frac{1}{2} y_{4u} (4 y_{2l}^2+16 y_{2u}^2+y_{3d}^2+y_{3u}^2 \nonumber \\
&& +2 (2
   y_{4d}^2+y_{4l}^2+5 y_{4u}^2+3 y_{5d}^2+y_{5l}^2+3 y_{5u}^2)) + a_u y_{4u} \,, \nonumber \\
16 \pi^2 \frac{d y_{5u}}{dt} &=& \frac{1}{2} (y_{5u} (y_{1d}^2+y_{1u}^2+6 y_{2d}^2+4 y_{2l}^2+14 y_{2u}^2+2 y_{3u}^2+6 y_{4d}^2+2 y_{4l}^2+6 y_{4u}^2+3 y_{5d}^2+2 y_{5l}^2) \nonumber \\
&& -4 y_{1d} y_{3u} y_{4d}+11 y_{5u}^3) + a_u y_{5u}\,, \nonumber \\
16 \pi^2 \frac{d y_{1d}}{dt} &=& \frac{1}{2} (15 y_{1d}^3+y_{1d} (4 y_{1l}^2+9 y_{1u}^2+6 y_{2d}^2+2 y_{2u}^2+6 y_{3d}^2+2 y_{3l}^2+6 y_{3u}^2+2 y_{4d}^2+y_{5d}^2+y_{5u}^2) \nonumber \\
&&-4 (2
   y_{1u} y_{2d} y_{2u}+y_{3u} y_{4d} y_{5u})) + a_d y_{1d} \,, \nonumber \\
16 \pi^2 \frac{d y_{2d}}{dt} &=& \frac{1}{2} (3 y_{1d}^2 y_{2d}-4 y_{1d} y_{1u} y_{2u}+y_{1u}^2 y_{2d}+18 y_{2d}^3+4 y_{2d} y_{2l}^2+14 y_{2d} y_{2u}^2+8 y_{2d} y_{4d}^2+2 y_{2d}
   y_{4l}^2 \nonumber \\
   &&+6 y_{2d} y_{4u}^2+7 y_{2d} y_{5d}^2+2 y_{2d} y_{5l}^2+3 y_{2d} y_{5u}^2-4 y_{2u} y_{4d} y_{4u}) + a_d y_{2d}\,, \nonumber \\
16 \pi^2 \frac{d y_{3d}}{dt} &=& 6 y_{1d}^2 y_{3d}+2 y_{1l}^2 y_{3d}+6 y_{1u}^2 y_{3d}-4 y_{1u} y_{4u} y_{5d}+\frac{9 y_{3d}^3}{2}+y_{3d} y_{3l}^2+\frac{3 y_{3d} y_{3u}^2}{2} \nonumber \\
&& +y_{3d}
   y_{4d}^2+y_{3d} y_{4u}^2+2 y_{3d} y_{5d}^2 + a_d y_{3d} \,, \nonumber \\
16 \pi^2 \frac{d y_{4d}}{dt} &=& y_{1d}^2 y_{4d}-2 y_{1d} y_{3u} y_{5u}+8 y_{2d}^2 y_{4d}-4 y_{2d} y_{2u} y_{4u}+\frac{1}{2} y_{4d} (4 y_{2l}^2+12 y_{2u}^2+y_{3d}^2+y_{3u}^2 \nonumber \\
&& +2 (5
   y_{4d}^2+y_{4l}^2+2 y_{4u}^2+3 y_{5d}^2+y_{5l}^2+3 y_{5u}^2)) + a_d y_{4d} \,, \nonumber \\
16 \pi^2 \frac{d y_{5d}}{dt} &=& \frac{1}{2} (y_{5d} (y_{1d}^2+14 y_{2d}^2+4 y_{2l}^2+6 y_{2u}^2+2 y_{3d}^2+6 y_{4d}^2+2 y_{4l}^2+6 y_{4u}^2+11 y_{5d}^2+2 y_{5l}^2+3 y_{5u}^2) \nonumber \\
&& +y_{1u}^2 y_{5d}-4 y_{1u} y_{3d} y_{4u}) + a_d y_{5d} \,, \nonumber \\
16 \pi^2 \frac{d y_{1l}}{dt} &=& \frac{1}{2}  y_{1l} (12 y_{1d}^2 + 7 y_{1l}^2 + 
   2 (6 y_{1u}^2 + 3 y_{2l}^2 + 3 y_{3d}^2 + y_{3l}^2 + 3 y_{3u}^2 + y_{4l}^2) + y_{5l}^2) + a_l y_{1l}\,, \nonumber \\
16 \pi^2 \frac{d y_{2l}}{dt} &=& \frac{1}{2} y_{2l} (3 y_{1l}^2 + 12 y_{2d}^2 + 10 y_{2l}^2 + 12 y_{2u}^2 + 6 y_{4d}^2 + 
   4 y_{4l}^2 + 6 y_{4u}^2 + 6 y_{5d}^2 + 3 y_{5l}^2 + 6 y_{5u}^2) + a_l y_{2l}\,, \nonumber \\
16 \pi^2 \frac{d y_{3l}}{dt} &=& \frac{1}{2} y_{3l} (12 y_{1d}^2 + 4 y_{1l}^2 + 12 y_{1u}^2 + 6 y_{3d}^2 + 5 y_{3l}^2 + 
   6 y_{3u}^2 + 2 y_{4l}^2 + 4 y_{5l}^2) + a_l y_{3l}\,, \nonumber \\
16 \pi^2 \frac{d y_{4l}}{dt} &=& \frac{1}{2} y_{4l} (2 y_{1l}^2 + 12 y_{2d}^2 + 8 y_{2l}^2 + 12 y_{2u}^2 + y{3l}^2 + 6 y_{4d}^2 + 
   6 y_{4l}^2 + 6 y_{4u}^2 + 6 y_{5d}^2 + 2 y_{5l}^2 + 6 y_{5u}^2) \nonumber \\
   &&+ a_l y_{4l}\,, \nonumber \\ 
   16 \pi^2 \frac{d y_{5l}}{dt} &=& \frac{1}{2} y_{5l} (y_{1l}^2 + 12 y_{2d}^2 + 6 y{2l}^2 + 12 y_{2u}^2 + 2 y_{3l}^2 + 6 y_{4d}^2 + 
   2 y_{4l}^2 + 6 y_{4u}^2 + 6 y_{5d}^2 + 7 y_{5l}^2 + 6 y_{5u}^2) \nonumber \\
   &&+ a_l y_{5l}\,.
\eea
With 
\bea
a_d &=& - 8 g_s^2 - \frac{9}{4} g^2 - \frac{5}{12} g'^2  \,, \nonumber \\
a_u &=& - 8 g_s^2 - \frac{9}{4} g^2 - \frac{17}{12} g'^2 \,,\nonumber \\
a_l &=& - \frac{9}{4} g^2 - \frac{15}{4} g'^2 \,.
\eea




\section{\\Couplings}
\label{appendix_couplings}
Below we show the interactions with the neutral $CP$-even scalars $h, H_1, H_2$ with the gauge bosons $V = W^{\pm},Z$:

\besub
\bea
g_{hVV} &=& \big(O_{11}s_\beta c_\gamma + O_{21}s_\beta s_\gamma
 + O_{31} c_\beta\big) \frac{n M_V^2}{v} \\
g_{H_1VV} &=& \big(O_{12}s_\beta c_\gamma + O_{22}s_\beta s_\gamma
 + O_{32} c_\beta\big)  \frac{n M_V^2}{v} \\
g_{H_2VV} &=& \big(O_{13}s_\beta c_\gamma + O_{23}s_\beta s_\gamma
 + O_{33} c_\beta\big)  \frac{n M_V^2}{v}
\eea
\eesub

Here $n = 2(1)$ for $W^{\pm}(Z)$.

Flavour-conserving couplings of $h$ with $u$-quarks
\besub
\bea
y_{huu} &=& O_{31} y_{1u} - O_{21} s_\gamma y_{2u}
- O_{11} c_\gamma y_{2u} \\
y_{hcc} &=& O_{31} y_{1u} + O_{21} s_\gamma y_{2u}
+ O_{11} c_\gamma y_{2u} \\
y_{htt} &=& \frac{O_{31}}{c_\beta} y_{3u}
\eea
\eesub

Flavour-violating couplings with $u$-quarks
\besub
\bea
y_{huc} &=& \frac{y_{2u}}{\sqrt{2}}\Big(-O_{21} c_\gamma
 + O_{11} s_\gamma\Big) \\
y_{hut} &=& \frac{y_{5u}}{2}\Big(O_{21} \sqrt{1 + s_\gamma}
- O_{11} \sqrt{1 - s_\gamma}) \\
y_{hct} &=& \frac{y_{5u}}{2}\Big(O_{21} \sqrt{1 - s_\gamma}
+ O_{11} \sqrt{1 + s_\gamma}) \\
y_{H_1uc} &=& \frac{y_{2u}}{\sqrt{2}}\Big(-O_{22} c_\gamma
 + O_{12} s_\gamma\Big) \\
y_{H_1ut} &=& \frac{y_{5u}}{2}\Big(O_{22} \sqrt{1 + s_\gamma}
- O_{12} \sqrt{1 - s_\gamma}) \\
y_{H_1ct} &=& \frac{y_{5u}}{2}\Big(O_{22} \sqrt{1 - s_\gamma}
+ O_{12} \sqrt{1 + s_\gamma}) \\
y_{H_2uc} &=& \frac{y_{2u}}{\sqrt{2}}\Big(-O_{23} c_\gamma
 + O_{13} s_\gamma\Big) \\
y_{H_2ut} &=& \frac{y_{5u}}{2}\Big(O_{23} \sqrt{1 + s_\gamma}
- O_{13} \sqrt{1 - s_\gamma}) \\
y_{H_2ct} &=& \frac{y_{5u}}{2}\Big(O_{23} \sqrt{1 - s_\gamma}
+ O_{13} \sqrt{1 + s_\gamma})
\eea
\eesub
Corresponding couplings for the down-sector can be obtained by the replacements $u \rightarrow d$, $c \rightarrow s$ and $t \rightarrow b$.

It is noted that the flavour-violating couplings of $A_1$($A_2$) are same as the corresponding ones of $H_1$($H_2$).

\bibliography{3HDM_ref}{}

\end{document}